\documentclass{jfm}
\usepackage{bm}
\usepackage{natbib}
\usepackage[dvips]{graphicx}
\usepackage{amsmath}

\newcommand{\secref}[1]{section \ref{#1}}
\newcommand{\secsand}[2]{sections \ref{#1} and \ref{#2}}
\newcommand{\apref}[1]{appendix \ref{#1}}
\newcommand{\exref}[1]{(\ref{#1})}
\newcommand{\exsdash}[2]{(\ref{#1}-\ref{#2})}

\renewcommand{\eqref}[1]{equation\ (\ref{#1})}
\newcommand{\eqsref}[1]{equations\ (\ref{#1})}
\newcommand{\eqsdash}[2]{equations\ (\ref{#1}-\ref{#2})}
\newcommand{\eqsand}[2]{equations\ (\ref{#1}) and (\ref{#2})}
\newcommand{\Eqref}[1]{Equation\ (\ref{#1})}

\newcommand{\Eqsand}[2]{Equations\ (\ref{#1}) and (\ref{#2})}
\newcommand{\figref}[1]{figure\ \ref{#1}}

\newcommand{\bea}{\begin{eqnarray}}
\newcommand{\eea}{\end{eqnarray}}
\newcommand{\beq}{\begin{equation}}
\newcommand{\eeq}{\end{equation}}
\newcommand{\lt}{\left}
\newcommand{\rt}{\right}
\newcommand{\bl}{\bigl}
\newcommand{\br}{\bigr}

\newcommand{\mbf}[1]{\bm{#1}}

\newcommand{\dd}{\partial}
\newcommand{\vdel}{\boldsymbol{\nabla}}
\newcommand{\vdperp}{\vdel_\perp}
\newcommand{\dperp}{\nabla_\perp}

\newcommand{\const}{\mathrm{const}}

\newcommand{\vr}{\mbf{r}}
\newcommand{\vk}{\mbf{k}}
\newcommand{\kperp}{k_\perp}
\newcommand{\vkperp}{\mbf{k}_\perp}
\newcommand{\kperpc}{k_{\perp c}}
\newcommand{\kpar}{k_\parallel}
\newcommand{\lpar}{l_\parallel}
\newcommand{\vperp}{v_\perp}
\newcommand{\vpar}{v_\parallel}
\newcommand{\uperp}{u_\perp}
\newcommand{\upar}{u_\parallel}
\newcommand{\vu}{\mbf{u}}
\newcommand{\vuperp}{\vu_\perp}
\newcommand{\vvort}{\delta\mbf{\omega}}
\newcommand{\operp}{\delta\omega_\perp}
\newcommand{\vomega}{\mbf{\omega}}
\newcommand{\vOmega}{\mbf{\Omega}}
\newcommand{\vz}{\mbf{\hat z}}
\newcommand{\vb}{\mbf{\hat b}}
\newcommand{\domega}{\delta\omega}
\newcommand{\urms}{u_{\rm rms}}

\newcommand{\vBperp}{\delta\mbf{B}_\perp}

\newcommand{\tnl}{\tau_{\rm NL}}

\newcommand{\Ro}{\mathrm{Ro}}

\newcommand{\eps}{\varepsilon}

\begin{document}

\title[Critical balance: towards a universal scaling conjecture]{Critical balance in magnetohydrodynamic, rotating and stratified turbulence: towards a universal scaling conjecture}
\author[S.\ V.\ Nazarenko and A.\ A.\ Schekochihin]{Sergei~V.~Nazarenko$^{1,3}$ and Alexander~A.~Schekochihin$^{2,3}$}
\affiliation{$^1$Mathematics Institute, University of Warwick, Coventry CV4 7AL, UK\\
$^2$Rudolf Peierls Centre for Theoretical Physics, University of Oxford, Oxford OX1 3NP, UK\\
$^3$Institut Henri Poincar\'e, Universit\'e Pierre et Marie Curie, 75231 Paris Cedex 5, France}
\date{22 April 2009; Published in {\em J.~Fluid Mech.}~{\bf 677}, 134 (2011)}

\maketitle

\begin{abstract}
It is proposed that critical balance --- a scale-by-scale balance between 
the linear propagation and nonlinear interaction time scales --- can be 
used as a universal scaling conjecture for determining the spectra of strong
turbulence in anisotropic wave systems. Magnetohydrodynamic (MHD), rotating and 
stratified turbulence are considered under this assumption and, in particular, 
a novel and experimentally testable energy cascade scenario and a set of scalings 
of the spectra are proposed for low-Rossby-number rotating turbulence. 
It is argued that in neutral fluids, 
the critically balanced anisotropic cascade provides a natural path from 
strong anisotropy at large scales to isotropic Kolmogorov turbulence at 
very small scales. It is also argued that the $\kperp^{-2}$ spectra seen in 
recent numerical simulations of low-Rossby-number rotating turbulence may be 
analogous to the $\kperp^{-3/2}$ spectra of the numerical MHD turbulence in 
the sense that they could be explained by assuming that fluctuations are 
polarised (aligned) approximately as inertial waves (Alfv\'en waves for MHD). 
\end{abstract}

\section{Introduction} 

Ability to support both linear waves and nonlinear interactions is 
ubiquitous in natural systems. Wave turbulence is, therefore, a very generic 
situation in such systems 
when dissipation coefficients are small and energy injected at some system-specific 
scale has to be dissipated at much smaller scales \citep{Zakharov92}. 
Theory of turbulence is concerned with 
the ways in which the energy is transferred from large (injection) to small (dissipation) scales 
and, consequently, with the structure of the fluctuations in the intervening scale range.

A common property of many such systems is the presence of some mean 
field that introduces a special direction. Examples are 
plasmas embedded in a mean magnetic field, 
rotating fluids and stably stratified fluids 
with a mean temperature or density gradient 
(in real systems usually in the direction of gravity). 
Both linear and nonlinear physics is affected by the mean field: 
turbulent fluctuations in such systems tend to display 
a high degree of anisotropy. The typical wave numbers parallel and 
perpendicular to the special direction associated with the mean field 
often satisfy $\kpar\ll\kperp$, while the wave dispersion 
relation is of the form 
\beq
\omega = \kpar v(\kperp).
\label{dr_gen}
\eeq
For Alfv\'en waves in magnetohydrodynamics (MHD), $v=v_A$, the Alfv\'en speed,
for inertial waves in 
rotating fluids, $v=2\Omega/\kperp$, where $\Omega$ is the rotation frequency.
Our arguments will apply directly to these two cases; 
in stratified turbulence, $\kpar\gg\kperp$, so the roles 
of $\kpar$ and $\kperp$ are reversed and certain adjustments 
to the general argument will be needed --- they are explained in 
\secref{sec_strat}. 
Note that low-frequency waves in magnetised plasmas generally satisfy the 
gyrokinetic dispersion relation, which is also of the form \exref{dr_gen} 
\citep{Howes06} and of which the Alfv\'en-wave dispersion relation is
a large-scale limiting case. 

The dispersion relation of the type \exref{dr_gen} implies that waves 
propagate primarily in the parallel direction: indeed, 
the parallel and perpendicular group velocities 
are $\vpar = v(\kperp)$ and 
$\vperp = (\kpar/\kperp) \kperp v'(\kperp)\ll\vpar$. 
If the nonlinearity is of the fluid type, $\vu\cdot\vdel\vu$, 
where $\vu$ is the fluid velocity, then
$\kpar\ll\kperp$ implies that nonlinear interactions 
are primarily perpendicular: $\vu\cdot\vdel\vu\simeq\vu_\perp\cdot\vdel_\perp\vu$,
so the nonlinear decorrelation time is given by
\beq
\tnl^{-1}\sim \kperp\uperp(\kperp),
\label{tnl_gen}
\eeq  
where $\uperp(\kperp)$ is the characteristic velocity fluctuation 
amplitude corresponding to the wave number $\kperp$
%see \secref{sec_align} 
%for an explanation of when \eqref{tnl_gen} is valid and what happens when it is not).
(this formula assumes that fluctuations are not polarised in any particular 
way that might reduce the nonlinear interactions; if one assumes 
they are, in fact, so polarised, the scaling theory presented below 
has to be modified as explained in \secref{sec_align}). 
Note that incompressibility $\vdel\cdot\vu=0$ and $\kpar\ll\kperp$ 
imply that the perpendicular motions are individually incompressible, 
$\vdperp\cdot\vuperp=0$ (see \apref{ap_reduced}). 

For anisotropic wave systems, Kolmogorov-style dimensional 
theory alone does not fix the scalings of the energy spectra. 
Indeed, assuming a local (in scale) energy cascade 
and hence a scale-independent energy flux~$\eps$,
\beq
\kperp E(\kperp)\sim \uperp^2 (\kperp) \sim \eps\tau(\kperp),
\label{E_gen}
\eeq
where $E(\kperp)$ is the one-dimensional perpendicular
energy spectrum and $\tau(\kperp)$ is the ``cascade time'' corresponding 
to the characteristic wave number $\kperp$. In the absence of waves or 
anisotropy, it is dimensionally inevitable that $\tau(k)\sim\tnl(k)$, 
whence $E(\kperp)\sim\eps^{2/3} k^{-5/3}$, the Kolmogorov spectrum. 
However, with waves and anisotropy, two additional 
dimensionless ratios arise: $\kpar/\kperp$ and 
$\omega\tnl\sim\kpar v/\kperp\uperp$, which measure the strength 
of the anisotropy and the relative time scales of the 
linear propagation and nonlinear interaction (equivalently,
the relative size of the fluid velocity and the wave phase speed). 
The spectrum can, as far as dimensional theory is concerned, 
be an arbitrary function of these two ratios, both of which 
can have some nontrivial scaling with $\kperp$. 

Clearly, an additional physical assumption is necessary to fix 
the scalings. In strong MHD turbulence, it is 
known as the {\em critical balance (CB)} and states that the characteristic 
linear and nonlinear times are approximately equal at all scales: 
$\omega\sim\tnl^{-1}$ \citep{Higdon84,GS95}. We propose that CB 
be adopted more generally as a universal scaling conjecture for 
anisotropic wave turbulence. 
This removes the dimensional ambiguity in determining the 
cascade time, so we may set $\tau\sim\tnl$ 
and recover the Kolmogorov spectrum, 
\beq
E(\kperp)\sim \eps^{2/3}\kperp^{-5/3}.
\label{cb_gen}
\eeq 
The CB itself, $\omega\sim\tnl^{-1}$, then gives us 
a relationship between the parallel and perpendicular scales: 
\beq
\kpar \sim \lt[\tnl(\kperp)v(\kperp)\rt]^{-1}
\sim \eps^{1/3}\lt[v(\kperp)\rt]^{-1}\kperp^{2/3}.
\label{aniso_gen}
\eeq
Note that while these scaling arguments suggest that 
in some appropriately defined sense the energy distribution 
in the $(\kperp,\kpar)$ plane will have a peak along the CB curve 
\exref{aniso_gen}, they do not tell us what the 
functional shape of this distribution is (the ``width'' of the peak). 
However, as we will see in \secref{sec_aniso}, \eqref{aniso_gen} 
is in fact sufficient to produce a testable quantitative prediction 
of the energy spectrum with $\kpar$. 

In what follows, we 
first give a general argument in favour of the idea of CB (\secref{sec_gen}) 
and then discuss three examples: MHD (and, more generally, plasma) turbulence, 
from whence these ideas originate (\secref{sec_MHD}), 
rotating turbulence, for which we propose a novel energy cascade 
scenario (\secref{sec_rot}), and stratified turbulence (\secref{sec_strat}). 
Note that in \secref{sec_align}, we propose the extension to the rotating 
turbulence of the concept of {\em polarisation alignment} (also originating from 
MHD turbulence; see \citealt{Boldyrev06}), which may help interpret 
the $\kperp^{-2}$ spectra reported in numerical simulations \citep{Mininni09,Thiele09}.  
The section on rotating turbulence is the main part of this paper, 
while the MHD and stratified cases are discussed only briefly to emphasise 
what appears to be universal nature of some of the scaling arguments involved. 
In \secref{sec_plasma} and in 
our concluding remarks (\secref{sec_conc}), we will also mention a few other 
examples of CB emerging as a general physical principle
in wave systems, including those that are different from the anisotropic 
wave type discussed here. 

\section{Why anisotropic turbulence is neither weak nor two-dimensional}
\label{sec_gen}

%Historically, turbulent wave systems have been approached 
%by noticing that 
In turbulent wave systems, 
if the fluctuation amplitudes at the injection 
scale are so small that $\omega\tnl\gg1$, the nonlinearity can be 
treated perturbatively and what is known as weak turbulence theory 
emerges as a controlled approximation \citep{Zakharov92}. 
In anisotropic wave systems, it typically predicts a turbulent 
cascade primarily in $\kperp$ (because the nonlinearity 
is primarily perpendicular), at constant $\kpar$ 
--- either exactly \citep[in MHD; see][]{Galtier00} 
or approximately \citep[in rotating turbulence; see][]{Galtier03}.
While the analytic calculations can be quite involved, 
the basic result can be recovered 
in a simple nonrigorous way. If $\omega\tnl\gg1$, nonlinear interactions between 
wave packets result in small perturbations of the 
amplitudes $\delta\uperp\sim (\omega\tnl)^{-1}\uperp$. 
These perturbations can be assumed to accumulate as a random walk 
and then the cascade time $\tau$ is by definition the time that 
it takes the cumulative perturbation to become comparable to the 
amplitude itself: $n^{1/2}\delta\uperp\sim\uperp$, where 
$n\sim\tau\omega$ is the number of interactions. This 
gives $\tau\sim\omega\tnl^2$ and, using \eqsdash{dr_gen}{E_gen}, we get 
the one-dimensional perpendicular energy spectrum
\beq
E(\kperp)\sim (\eps\kpar)^{1/2} \lt[v(\kperp)\rt]^{1/2}\kperp^{-2}.
\label{weak_gen}
\eeq
We stress that since we have {\em assumed} that 
there is no parallel cascade, the energy injection rate $\eps$
can be an arbitrary function of $\kpar$.
Thus, in \eqref{weak_gen} and in all other subsequent developments pertaining 
to weak turbulence, $\kpar$ is a {\em parameter} --- it is the characteristic 
parallel wave number 
at which energy is injected. For simplicity, one may assume that the
injection is isotropic and so $\kpar\sim k_0$, the energy-injection wave number
that will appear in \secsand{sec_MHD}{sec_rot}.  
Note that in other anisotropic wave systems there can be a cascade in the
parallel direction. The weak turbulence spectra for some such systems 
(historically the first example of anisotropic wave spectra) 
were found by \citet{Kuznetsov72}. The parallel energy transfer in weak 
rotating turbulence (ignored by us) is discussed in great detail by \citet{Bellet06}.

Using \eqsref{weak_gen} and \exsdash{dr_gen}{E_gen}, it is easy to work out
the condition under which the weak turbulence approximation is valid: 
$\omega\tnl\sim\kpar v(\kperp)\eps^{-1/3}\kperp^{-2/3}\gg1$. 
Unless $v(\kperp)$ increases sufficiently fast with 
$\kperp$, the nonlinearity becomes stronger with increasing 
$\kperp$ compared to the linear propagation and 
the weak turbulence condition is broken at 
$\kperp$ given by \eqref{aniso_gen}.  
Thus, the weak turbulence cascade drives itself into 
a critically balanced state (see \citealt{Weak} for a somewhat 
less conventional but conceptually perhaps more convincing argument to 
this effect). 

The opposite limit %, motivated by the anisotropy, 
is a pure two-dimensional (2D) state: 
$\kpar$ is assumed so small that 
$\omega\tnl\ll1$ and wave propagation is neglected. 
As generically happens in 2D, the energy cascade should then be 
inverse, from larger to smaller $\kperp$. As $\kperp$ decreases, 
$\tnl$ becomes longer, so 
the 2D approximation, $\omega\tnl\ll1$, is eventually broken 
and CB is reached, at which point the turbulence is 
again three-dimensional (3D). Thus, both from the weak-turbulence limit
(small amplitudes) and the 2D limit, the turbulence naturally evolves 
towards a state of CB. 
%This situation is illustrated in \figref{fig_cascade}.

There exists another argument, which is independent 
of the assumption of inverse cascade and suggests that 
2D motions are fundamentally unstable. 
Consider two perpendicular planes separated by some distance. 
The motions in each plane will decorrelate on the time scale $\tnl$. 
%In a pure 2D state, the two planes have to decorrelate in exactly the same 
%way, otherwise fluctuations will develop a characteristic 
%$\kpar$ corresponding to the distance between the planes. 
In the parallel direction, information is transmitted by waves, 
so perfect correlation between the two planes required for a pure 2D state 
can only be sustained if a wave can propagate between them 
in a time shorter than $\tnl$. Thus, for any given $\kperp$, 
there will be some parallel distance, $\kpar^{-1}$, given by the CB relation 
\exref{aniso_gen}, beyond which the motions will decorrelate and become 3D. 
Thus, an initially 2D perturbation will tend to a state of CB
(this argument was suggested to us by S.~C.~Cowley, 2004).
This process can be interpreted as an instability of the 2D motions 
with respect to Cherenkov-type emission of waves. 

\section{MHD (Alfv\'enic) turbulence} 
\label{sec_MHD}

The ideas laid out above in a general 
form originate from considerations of MHD turbulence. 
In MHD, \eqref{dr_gen} describes Alfv\'en waves 
with $v=v_A=B_0/\sqrt{4\pi\rho}$, where $B_0$ is the mean magnetic field 
and $\rho$ the density of the conducting fluid. 
Small anisotropic fluctuations in such a turbulence 
are Alfv\'enic, $\uperp\sim\delta B_\perp/\sqrt{4\pi\rho}$, 
where $\delta B_\perp$ is the perpendicular perturbation 
of the magnetic field (mathematically this statement can be formalised 
in terms of the Reduced MHD equations; see the end of \apref{ap_reduced}). 

Both weak-turbulence \citep{Galtier00} and 
2D \citep{Montgomery81} theories for MHD turbulence have been proposed. 
By the general arguments given above, both will naturally evolve towards 
a CB state, $\kpar v_A\sim\kperp\uperp$, with 
a Kolmogorov spectrum \exref{cb_gen} and a scale-dependent 
anisotropy given by \eqref{aniso_gen}: 
\beq
\kpar\sim \eps^{1/3} v_A^{-1}\kperp^{2/3}. 
\label{aniso_MHD}
\eeq
Note that as $\kperp$ increases, 
the turbulence becomes more anisotropic ($\kpar/\kperp$ decreases).

If the turbulence is weak at the injection scale, its spectrum is 
expected to be [see \eqref{weak_gen}] 
\beq
E(\kperp)\sim (\eps\kpar v_A)^{1/2}\kperp^{-2}
\label{weak_MHD}
\eeq 
and there is no cascade in $\kpar$ \citep{Galtier00}. 
From \eqref{weak_MHD}, $\urms\sim (\eps v_A/k_0)^{1/4}$, 
where $k_0$ is the wave number of energy injection (assumed isotropic), so 
$\eps\sim M_A^4 v_A^3 k_0$, where $M_A = \urms/v_A\ll1$ is the Alfv\'enic Mach number. 
Then the wave number at which weak turbulence breaks down
and the critically balanced cascade begins is, from \eqref{aniso_MHD}, 
\beq
\kperpc \sim \eps^{-1/2} (\kpar v_A)^{3/2}\sim k_0 M_A^{-2}.
\label{kperpc_MHD}
\eeq

The anisotropy of Alfv\'enic turbulence 
and even \eqref{aniso_MHD} appear to have been 
confirmed by numerical simulations \citep{Cho00,Maron01} 
and solar wind measurements \citep{Horbury08,Podesta09,Wicks10,ChenMallet11}
--- this will be discussed further in \secref{sec_aniso}. 
The weak turbulence spectral scaling \exref{weak_MHD} has also 
been checked numerically \citep{Perez08}.  
The precise nature of the scaling of the energy spectrum in the CB regime 
remains somewhat mysterious: while the solar wind measurements support 
$\kperp^{-5/3}$ \citep[e.g.,][]{Bale05,Horbury08,Sahraoui09,Wicks10,ChenMallet11}, 
numerical simulations give spectra much closer 
to $\kperp^{-3/2}$ \citep{Maron01,Mason08} 
--- a modified critical balance argument proposed 
by \citet{Boldyrev06} to explain these results will be discussed 
in \secref{sec_align}, where we will show how it can be adapted 
to the case of rotating turbulence. 

\subsection{Plasma turbulence}
\label{sec_plasma}

Beyond the MHD approximation, the gyrokinetic dispersion relation 
for low-frequency waves in magnetised plasmas is also of the 
form \exref{dr_gen} \citep{Howes06}. The general idea of a critically 
balanced cascade can be extended to various types of gyrokinetic turbulence, 
e.g., for plasma turbulence below the ion Larmor scale \citep{Cho04,Tome}.
%For example, for plasma turbulence 
%at scales below the ion Larmor radius, which can be described by 
%the so-called Electron Reduced MHD approximation \cite{Tome}, 
%numerical confirmation of a critically balanced cascade 
%of kinetic Alfv\'en waves 
%\footnote{For kinetic Alfv\'en waves (KAW), 
%\eqref{dr_gen} holds with $v(\kperp)\propto v_A\kperp\rho_i$, where 
%$\rho_i$ is the ion Larmor radius. The rest of the general 
%argument presented above works for the KAW turbulence 
%with some adjustments due to a different type of nonlinearity 
%in plasmas at sub-Larmor scales: 
%\eqref{tnl_gen} becomes $\tnl^{-1}\sim \kperp^2\rho_i\bperp$, 
%where $\bperp$ is the magnetic-field fluctuation amplitude; 
%in \eqref{E_gen}, $\uperp$ has to be replaced by $\bperp$; 
%then in the critically balanced state, 
%$E(\kperp)\sim (\eps/\rho_i)^{2/3}\kperp^{-7/3}$ 
%and instead of \eqref{aniso_gen}, we get 
%$\kpar\sim (\eps/\rho_i)^{1/3}v_A^{-1}\kperp^{1/3}$. 
%Further details of this and similar arguments for kinetic plasma 
%turbulence can be found in \cite{Tome}, but here the important 
%point is that the argument proposed in this paper easily 
%generalises to plasma systems as long as the cascading quantity 
%(not always kinetic energy) and the type of nonlinearity (not always 
%$\vu\cdot\vdel\vu$) are correctly identified.}
%has been achieved \cite{Cho04,Cho09} and spectra measured in the 
%solar wind are consistent with this prediction \cite{Sahraoui09}.
Further details can be found in \citet{Tome} (Sec.~7); the important 
point to keep in mind is that generalising the argument proposed in the 
present paper to plasma systems requires correctly identifying the cascading 
quantity (not always kinetic energy) and the type of nonlinearity 
(not always $\vu\cdot\vdel\vu$).

In the case of the turbulence of 
kinetic Alfv\'en waves (dispersive waves 
that replace the MHD Alfv\'en waves below the ion Larmor scale), 
the scaling predictions resulting 
from the application of the CB conjecture to their 
dispersion relation (also of the form \exref{dr_gen})
have been confirmed numerically \citep{Biskamp99,Cho04,Cho09}; the 
sub-Larmor-scale spectra and structure functions measured in the 
solar wind also appear to be consistent with the CB prediction 
\citep{Sahraoui09,Chen10}. 
This is the first example of confirmed applicability of the CB conjecture 
beyond its original target of MHD turbulence. 

\section{Rotating turbulence} 
\label{sec_rot}

Are magnetic anisotropy and magnetised plasma waves a special case or 
can CB be adopted as a universal scaling 
conjecture? A key test of universality would be to find a critically 
balanced cascade in a purely hydrodynamic setting. 
We propose the following scenario for the rotating turbulence. 

\subsection{Critically balanced rotating turbulence and restoration of isotropy}
\label{sec_unaligned} 

The dispersion relation for inertial waves in a rotating 
incompressible fluid is $\omega = 2\Omega \kpar/k$.  
Suppose that the energy is injected isotropically 
at some characteristic wave number $k_0$ 
and the Rossby number $\Ro = \urms k_0/\Omega\ll1$, i.e.,  
the amplitudes at the injection scale are so low that 
turbulence is weak. Then the energy cascade will 
proceed anisotropically: let us assume for maximum simplicity 
that the parallel cascade is negligible, 
i.e., $\kpar$ stays of the order of $k_0$ while energy moves 
towards larger $\kperp$ \citep{Galtier03}. 
When $\kperp\gg\kpar$, the dispersion relation takes 
the form \exref{dr_gen} with $v(\kperp)=2\Omega/\kperp$; the weak 
turbulence spectrum is then given by \eqref{weak_gen} 
[analogous to \eqref{weak_MHD}]: 
\beq
E(\kperp)\sim (\eps k_0\Omega)^{1/2}\kperp^{-5/2}.
\label{weak_rot}
\eeq
Note that this implies $\urms\sim(\eps\Omega)^{1/4}k_0^{-1/2}$ 
and so $\eps\sim \Ro^4\Omega^3 k_0^{-2}$. 
As $\kperp$ increases, the nonlinearity becomes stronger 
and CB is 
reached at a critical $\kperp$ that can inferred from 
\eqref{aniso_gen} by setting $\kpar\sim k_0$ [analogous to \eqref{kperpc_MHD}]: 
\beq
\kperpc\sim \eps^{-1/5}(\kpar\Omega)^{3/5} \sim k_0\Ro^{-4/5}. 
\label{kperpc_rot}
\eeq
For $\kperp>\kperpc$, the turbulence is no longer weak, but it is 
still anisotropic and the cascade is critically balanced: 
the spectrum is given by \eqref{cb_gen}, while 
\eqref{aniso_gen} becomes 
\beq
\kpar \sim \eps^{1/3}\Omega^{-1}\kperp^{5/3} %\sim k_0 (\kperp/\kperpc)^{5/3}. 
\label{aniso_rot}
\eeq
\citep[cf.][]{Galtier05}.  
This relation is qualitatively different from the MHD case 
[\eqref{aniso_MHD}] in that 
the fluctuations become less, rather than more, anisotropic as 
$\kperp$ increases. Isotropy is reached 
when $\kperp\sim\kpar\sim k_i$, where 
\beq
k_i \sim \eps^{-1/2}\Omega^{3/2} \sim k_0 \Ro^{-2}
\label{ki_rot}
\eeq
\citep[cf.][]{Zeman94}. 
At this wave number, the velocity is $u(k_i)\sim (\eps/\Omega)^{1/2}$ 
(using \eqsand{E_gen}{cb_gen})
and so the Rossby number at the corresponding scale is 
$k_i u(k_i)/\Omega\sim 1$. Therefore, at $k > k_i$, rotation 
is irrelevant and turbulence is of the familiar isotropic 
Kolmogorov type, with $E(k)\sim \eps^{2/3}k^{-5/3}$. 
This is, of course, the physically inevitable outcome because 
unlike in the case of magnetised turbulence, which can feel the 
mean magnetic field at any scale, the hydrodynamic 
turbulence cannot feel the mean rotation rate if the local 
Rossby numbers are large. It is reassuring that the critically 
balanced cascade predicted by the general argument
proposed here has naturally led to this Kolmogorov state. 

\begin{figure}
\centering
\includegraphics[width=8cm]{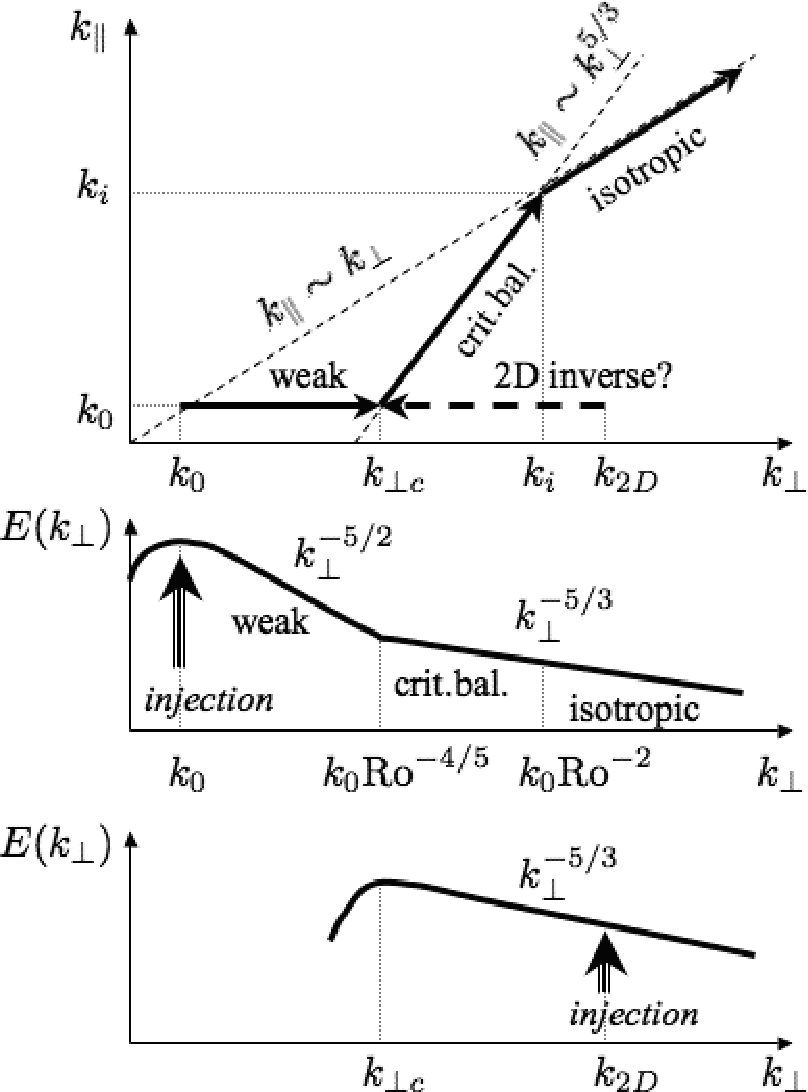}
\caption{A sketch of cascade path 
and spectra for the rotating turbulence: both 
the case of injection at $\kperp=\kpar=k_0$ 
(and $\Ro\ll1$) and that at $\kpar\ll\kperp=k_{\rm 2D}$ are shown
(see \secsand{sec_unaligned}{sec_inv_casc}). 
In the case of polarisation alignment, scalings shown in 
the two upper panels should be modified as explained in \secref{sec_align}.
The absence of the parallel cascade in the weak regime is a simplifying 
assumption, as acknowledged at the start of \secref{sec_unaligned}.}
%\Caption{Cascade path and spectra for the rotating turbulence
%with energy injection at $\kperp=\kpar=k_0$ and $\Ro\ll1$.}
\label{fig_rot} 
\end{figure}

The cascade path and the resulting spectrum are 
sketched in \figref{fig_rot}. We have illustrated the case discussed 
above, where energy is injected isotropically and in the weak 
turbulence regime. More generally, we expect that energy injected at 
a given $(\kperp,\kpar)$ will travel towards the CB path 
[\eqref{aniso_rot}] followed by isotropic Kolmogorov cascade
(as, e.g., shown in \figref{fig_rot} for the case of quasi-2D injection, 
discussed in \secref{sec_inv_casc}). Obviously, if the energy is injected 
at $k_0>k_i$, i.e., if $\Ro>1$ at the injection scale, the cascade will 
start and remain isotropic because rotation can be ignored.
Note that both the relationship \exref{aniso_rot} between the parallel and 
perpendicular scales and the wavenumber \exref{ki_rot} of the transition to 
isotropy depend only on $\eps$ and $\Omega$, but not on 
on the injection scale(s). Thus, the CB-to-isotropy path represents 
the ``natural'' state of rotating turbulence --- probably applicable 
to the decaying case as well as the forced one. Thus, 
we suspect that \eqref{aniso_rot} should prove to be a good prediction for the 
relationship between the parallel and perpendicular correlation lengths 
in the columnar vortical structures observed in experiments \citep[e.g.,][]{Davidson06}.

Our predictions of the spectral slopes \exref{weak_rot} and \exref{cb_gen} 
and the scaling of the transition wave number 
\exref{kperpc_rot} with $\Ro$ are experimentally and numerically verifiable
(but see \secref{sec_align} for possible alternative scalings). 
A transition from anisotropic to isotropic turbulence at 
the local $\Ro\sim1$ [corresponding to the wave number $k_i$, \eqref{ki_rot}] 
appears to have been observed first by \citet{Jacquin90}. 
A change of spectral slope from $\sim-2.2$ (perhaps consistent with $-5/2$) 
to $\sim-5/3$ [corresponding, in our theory, to the critical wave number $\kperpc$, 
\eqref{kperpc_rot}] 
may have been observed in rotating turbulence experiments with small 
initial $\Ro$ (Fig.~5b in \citealt{Morize05}), although it is premature 
to say if these results are definitely related to the 
transition to CB or merely to instrumental 
noise at high wave numbers. What does seem to be known definitely 
is that rotating turbulence has a clear tendency to a state where 
the local effective value of the Rossby number is $\sim1$, i.e., 
the linear and nonlinear time scales are comparable and both linear 
and nonlinear dynamics are manifestly present \citep{Davidson06}. 
This is consistent with the transition to CB that we are proposing 
and the nontrivial prediction is that whereas 
the spectrum is Kolmogorov for wave numbers above this transition 
$\kperp>\kperpc$, 
the turbulence remains anisotropic up to the second transition wave 
number $k_i$. 

In designing or interpreting both laboratory and numerical experiments 
to test our predictions, one has to be mindful of 
the following caveat. In order for the full cascade path sketched 
in \figref{fig_rot} to be realised, 
the system domain must be large enough to accommodate 
both the parallel and perpendicular scales implied by the CB state. 
If it is not, this will impose infrared cutoffs in the $(\kperp,\kpar)$ 
--- these can restrict energy flows, possibly leading to 2D effects such 
as inverse cascades and (in periodic numerical simulations) finite-box effects 
such as formation of a $\kpar=0$ condensate (see further discussion in 
\secref{sec_num}). 

Another important caveat concerns the absence (or negligibility) of the parallel cascade
in the weak-turbulence regime. Unlike in the case of MHD turbulence, 
for rotating turbulence this is not an exact result, 
but an assumption \citep{Galtier03} --- and possibly a gross 
simplification of a fairly complicated precise situation \citep{Bellet06}.
We have made this simplification because the detailed properties 
of the weak-turbulence regime are less important in our view than its 
general tendency towards a strongly nonlinear CB 
state, in which the exact form of the resonant manifold in 
the wavenumber space is irrelevant. 
%Note that there is some numerical support for the 
%assumption that the energy transfer in $\kpar$ is strongly quenched in 
%rotating turbulence \citep{Mininni09}.

\subsection{Local scale-dependent anisotropy} 
\label{sec_aniso}

Quantitatively checking \eqref{aniso_rot} is nontrivial. 
One possibility is to measure the energy spectrum 
with respect to parallel wave numbers: by 
definition, $\kpar E(\kpar)\sim\uperp^2\sim \kperp E(\kperp)$, 
so, using \eqsand{cb_gen}{aniso_rot}, we find a distinctive scaling:
\beq
E(\kpar) \sim \eps^{4/5}\Omega^{-2/5}\kpar^{-7/5},
\quad k_0 < \kpar < k_i, 
\quad \kperpc < \kperp < k_i
\label{Epar_rot}
\eeq
(note that this is only valid in the CB regime). 
There is a similar result for Alfv\'en-wave turbulence
based on \eqref{aniso_MHD}: 
\beq
E(\kpar)\sim\eps v_A^{-1}\kpar^{-2},
\quad \kpar > k_0, 
\quad \kperp > \kperpc
\label{Epar_MHD}
\eeq 
which has recently been corroborated by the 
solar wind measurements \citep{Horbury08,Podesta09,Wicks10,ChenMallet11}. 

It is from the MHD experience that one learns about an important 
subtlety in the definition of $\kpar$ or, more precisely, of the 
parallel scale $\lpar\sim\kpar^{-1}$ [in practice, 
scalings are usually extracted via structure functions 
rather than spectra: $\delta\uperp^2(l) \sim l^{-1}E(l^{-1})$]. 
In order for the scale-dependent anisotropy to become apparent, 
$\lpar$ had to be defined with respect to the ``local mean field,'' 
i.e., the global mean magnetic field plus its perturbations at all 
scales larger than the one we are interested in \citep{Cho00,Maron01,Horbury08,ChenMallet11}. 
Physically, this is because an Alfv\'enic perturbation can only ``see'' 
the local field, not the globally averaged one. Mathematically, 
measuring $\lpar$ with respect to the global mean field would 
only capture the anisotropy of the largest-scale 
perturbation, while for all smaller-scale ones, such globally 
defined $\lpar$ ``slips off'' one field line to a neighbouring 
one and effectively picks up perpendicular variation rather than the 
parallel one. 

Similarly, for rotating turbulence, we anticipate that 
some scheme might have to be devised to measure parallel 
correlations along the local mean vorticity direction rather 
than along the global rotation axis.
Indeed, it is physically 
intuitive that the inertial waves would propagate along the total 
vorticity $\vomega=2\vOmega + \vvort$, where $\vvort=\vdel\times\vu$
(see further discussion in \apref{ap_reduced}). 
In the CB regime, this deviation, while significant 
for measuring $\kpar$, is small: 
$\domega/\Omega\sim ku/\Omega\sim u/v\sim \kpar/\kperp\ll1$; 
once isotropy is restored, $\domega/\Omega\sim1$ (so inertial 
waves no longer have a definite direction of propagation).
%Note that, unlike in MHD, measuring $\kpar$ with respect to the global 
%rotation axis would produce a steeper scaling (-5/3 instead of -7/5) 
%and thus overestimate the anisotropy.
 
A practical method for measuring parallel correlations might be inspired by 
the wavelet \citep{Horbury08,Podesta09} or structure-function technique 
\citep{Chen10,ChenMallet11} used for the solar wind. 
Their method of measuring parallel spectra could in fact be taken as 
a good definition of what $\kpar$ precisely means.

\subsection{Is inverse cascade possible in rotating turbulence?} 
\label{sec_inv_casc}

Another interesting experimental possibility would be to stir 
the turbulence in a 2D way and find out whether it will 
develop an inverse cascade, bringing it first to the CB 
state and then to the isotropic Kolmogorov state. This possibility 
depends on the inverse cascade proceeding at a rate larger 
than the 2D structures radiating inertial waves, 
with energy thus directly transferred into the CB state
(see the argument at the end of \secref{sec_gen} regarding the instability 
of a 2D state).
  
A putative inverse cascade followed by a direct critically 
balanced cascade is sketched in \figref{fig_rot}. 
Suppose the energy is injected at $\kperp=k_{\rm 2D}\gg \kpar$.
%(see \figref{fig_rot}). 
The inverse energy cascade, if it is sustainable, 
will give rise to the spectrum \exref{cb_gen} for $\kperp<k_{\rm 2D}$. 
This will extend, presumably at constant $\kpar$, to 
$\kperp\sim\kperpc$, given by the first expression in \eqref{kperpc_rot}. 
At this point the turbulence is again 3D and the cascade should 
``turn around'' and follow the CB path as before. 
Interestingly, the net perpendicular energy flux 
(integrated over $\kpar$) is zero for $\kperp < k_{\rm 2D}$, 
although the spectrum is $\kperp^{-5/3}$ 
extending to wave numbers both larger and smaller than 
$k_{\rm 2D}$, with the infrared cutoff given by $\kperpc$. 
Since the velocity at $\kperpc$ is 
$\urms\sim\eps^{2/5}(\kpar\Omega)^{-1/5}$, we have 
$\kperpc\sim (\kpar\Omega/\urms)^{1/2}$, where $\kpar$ is 
the parallel wave number at which the energy was injected. 
Note that the cascade reversal is a nontrivial consequence 
of anisotropy; in isotropic turbulence, zero flux would imply 
thermodynamic energy equipartition, $E(k)\propto k^{2}$. 

Note that, as discussed at the end of 
\secref{sec_unaligned}, a 2D inverse cascade can also 
occur in a geometrically constrained situation where the system domain 
restricts the cascade path and makes the turbulence effectively 2D: 
e.g., if the infrared cutoff in $\kperp$ lies to the right of the 
CB line in \figref{fig_rot} (then the CB state cannot be reached
and no cascade reversal is possible). Mathematically this means that 
the $\dd/\dd z$ terms in \eqsand{Phi_eq}{upar_eq} are negligible
--- without them, the perpendicular velocity decouples 
from the parallel one, the latter becomes a passive tracer and the 
former a 2D velocity field unaware of the rotation.

\subsection{Numerical evidence and finite-box effects}
\label{sec_num}

There is a large body of literature on numerical simulations of 
rotating turbulence. The two most recent and best resolved 
numerical studies are by \citet{Mininni09} and \citet{Thiele09}. 
We refer the reader to these papers for a comprehensive list of 
references to previous numerical work, which we will not replicate here. 
Let us discuss the results, which appear to be consistent 
in many independent investigations. 

Relating numerical evidence 
to scaling theories like the one proposed above 
is far from straightforward because simulations are typically 
done in periodic boxes and we have not discussed the effects of finite 
dimensions of the containing volume on wave turbulence. For MHD turbulence,
it was shown by \citet{Nazarenko07} that a finite box size 
along the mean magnetic field can lead to suppression of nonlinear interactions
between modes with different $\kpar$. 
This results in a qualitatively different evolution of the 2D 
non-propagating $\kpar=0$ mode and the wave modes with finite $\kpar$
\citep[see also][]{Bourouiba08,Boldyrev09weak,Weak}. 
Similar effects are probably operative in the numerical simulations of rotating
turbulence, %\citep[cf.][]{Cambon04,Bellet06}, 
especially in relatively shallow boxes (because of the 
anisotropy, even cubic boxes are effectively shallow --- this is well 
known in MHD turbulence, where simulations are routinely done in long boxes;
see, e.g., \citealt{Maron01,Mason08}).
Both \citet{Mininni09} and \citet{Thiele09} 
(as well as many previous publications, e.g., \citealt{Smith99}) 
report significant accumulation of energy in the $\kpar=0$ modes, 
via a nonlocal inverse cascade. The $\kpar=0$ modes also have a 
different spectrum than the finite-$\kpar$ modes. 
Note that in the local inverse cascade scenario of
\secref{sec_inv_casc}, we envisioned energy injection at very small, 
but finite $\kpar$ and did not consider the dynamics of 
the exact $\kpar=0$ modes (whose existence is particular to numerical boxes). 

Another feature of the numerical simulations 
where the Rossby numbers associated with the forcing scale 
are low (the case we are considering in this paper) 
is what appears to be a robust $\kperp^{-2}$ scaling of the energy 
spectrum (see papers cited above and references therein). 
Is this a contradiction with the scaling predictions of 
\secref{sec_unaligned}? It must be stressed here that the appearance 
of the $\kperp^{-2}$ spectrum cannot be explained by theories that assume 
weak {\em isotropic} turbulence and infer a $k^{-2}$ spectrum 
\citep{Dubrulle92,Zeman94,Zhou95,Canuto97}, because numerical evidence 
appears clear that the turbulence is not isotropic. 
A similar problem was encountered 
in interpreting numerical simulations of MHD turbulence, 
which consistently give $E\sim\kperp^{-3/2}$ \citep{Maron01,Mason08}, 
rather than $\kperp^{-5/3}$ that followed from the scaling 
arguments of \secref{sec_MHD}. There again, the $k^{-3/2}$ spectrum that
MHD turbulence would have if it were weak and isotropic \citep{Iroshnikov63,Kraichnan65} 
is not relevant because MHD turbulence in these simulations is provably anisotropic. 
To resolve this problem, \citet{Boldyrev06} proposed a modification of the 
CB argument based on the idea that nonlinearity 
is weakened in a scale-dependent fashion if the fluctuating 
fields align in a certain way. 
It turns out that a similar modification 
can be constructed for rotating turbulence and 
yields a spectrum that agrees with numerical evidence. 

\subsection{Polarisation alignment}
\label{sec_align}

The estimate \exref{tnl_gen} for the nonlinear decorrelation 
time was correct subject to assuming implicitly that fluctuations 
are not polarised in any particular way that might weaken 
the nonlinearity, i.e., that the direction of $\vuperp$ 
decorrelates over the same scale as its amplitude. 
Let us consider what happens if we suppose instead that a typical 
turbulent fluctuation is three-dimensionally anisotropic with characteristic 
wave numbers $k_x\gg k_y\gg k_z\equiv\kpar$, where $x$ is the direction 
of maximum gradients remaining approximately the same throughout 
the fluctuation and $z$ the direction of the propagation 
of the inertial waves. Then, since in a system with $\kpar\gg\kperp$ 
the perpendicular velocity is individually incompressible, 
$\vdperp\cdot\vuperp=0$, we have $u_x \sim (k_y/k_x) u_y \ll u_y$. 
Note that if we took $k_y=0$, we would simply have a monochromatic 
inertial wave, which, as it is easy to show, is an exact nonlinear 
solution (see \apref{ap_reduced}). However, if a wave packet of 
such waves is introduced, there would be nonlinear interaction 
and we are now attempting to determine how much the inertial-wave-like 
polarisation of fluctuations can be preserved in a strongly turbulent 
nonlinear state. 

To estimate the nonlinear decorrelation time, we now replace 
\eqref{tnl_gen} with 
\beq
\label{tnl_align}
\tnl^{-1}\sim k_y u_y \sim \kperp\uperp(\kperp)\theta(\kperp),
\eeq 
where $\kperp\sim k_x$, $\uperp\sim u_y$ and 
$\theta\sim k_y/k_x \sim u_x/u_y\ll1$ is the velocity 
angle responsible for weakening the nonlinearity
($\theta=0$ would correspond to an inertial wave). 
To determine this angle, we need a physical hypothesis 
about the degree to which the inertial-wave polarisation of the 
velocity field is preserved across a typical turbulent fluctuation.
We argued in \secref{sec_aniso} that, physically speaking, inertial 
waves should propagate along the local mean vorticity direction rather 
than the global rotation axis and so the direction of anisotropy 
is scale-dependent. Thus, all directions within a fluctuation 
are determined to within an angular uncertainty 
$\delta\theta\sim\operp/\Omega$ set by the local value of the 
perpendicular vorticity fluctuation. One might then postulate 
that $\theta\sim\delta\theta$, i.e., that there is a tendency 
to preserve the inertial-wave polarisation to the maximal possible 
degree ({\em polarisation alignment} conjecture). Then 
\beq
\label{theta_align}
\theta\sim\frac{\operp}{\Omega}
\sim \frac{k_x\upar}{\Omega} \sim \frac{u_y}{v},
%\sim \frac{\kpar}{k_y},
\eeq
where we have taken $\upar\sim u_y$ (as in an inertial wave)
and used $v\sim\Omega/k_x$. Therefore, \eqref{tnl_align} becomes
\beq
\label{tnl_align2}
\tnl^{-1}\sim \kperp\lt[\uperp(\kperp)\rt]^2\lt[v(\kperp)\rt]^{-1}.
\eeq
Note that if we use the CB conjecture, $\tnl^{-1}\sim \kpar v$, 
and the fact that $\theta\sim k_x/k_y$ by definition, 
we find from \eqsand{tnl_align}{theta_align} that
$\theta\sim\kpar/k_y\sim(\kpar/k_x)^{1/2}$, 
$k_y \sim (\kpar k_x)^{1/2}$, and \eqref{tnl_align2} can be rewritten 
as $\tnl^{-1}\sim (\kperp\kpar)^{1/2}\uperp(\kperp)$.

Combining \eqsand{tnl_align2}{E_gen} and using CB, 
$\tau\sim\tnl\sim(\kpar v)^{-1}$, we~find 
\bea
\label{E_rot_align}
E(\kperp)\sim\lt[\eps v(\kperp)\rt]^{1/2}\kperp^{-3/2}
\sim(\eps\Omega)^{1/2}\kperp^{-2},\\
\kpar\sim\eps^{1/2}\lt[v(\kperp)\rt]^{-3/2}\kperp^{1/2}
\sim\eps^{1/2}\Omega^{-3/2}\kperp^2.
\label{aniso_rot_align}
\eea
These formulae replace \eqsand{cb_gen}{aniso_rot} for rotating 
turbulence with polarisation alignment. \Eqref{aniso_rot_align} 
implies that isotropy is again restored at the  
wave number $k_i$ given by \eqref{ki_rot},
while the transition wave number from weak to critically balanced turbulence is, 
instead of \eqref{kperpc_rot}, determined by
\beq
\kperpc\sim \eps^{-1/4}\kpar^{1/2}\Omega^{3/4} \sim k_0\Ro^{-1}.
\label{kperpc_rot_align}
\eeq
The sketch of the cascade path in \figref{fig_rot} is still valid, 
but with the new scalings for the CB regime (so at $\kperp=\kperpc$, 
the spectral slope now changes from $-5/2$ to $-2$ and at $\kperp=k_i$, 
from $-2$ to $-5/3$; for $\kperpc<\kperp<k_i$, $\kpar\propto\kperp^2$). 
The parallel spectrum is derived as in \secref{sec_aniso} 
and so \eqref{Epar_rot} is replaced by
\beq
E(\kpar) \sim \eps\lt[v(\kperp)\rt]^{-1}\kpar^{-2}
\sim \eps^{3/4}\Omega^{-1/4}\kpar^{-3/2}.
\label{Epar_rot_aniso}
\eeq

Interestingly, 
\citet{Dubrulle92} argue that a transition from $k^{-2}$ to $k^{-5/3}$ 
scaling is observed in the spectrum of motions in the galactic disk
and might be attributable to a transition from rotating to standard 
isotropic Kolmogorov turbulence (although in their theory, the rotating 
turbulence is isotropic and weak, so the origin of their $k^{-2}$ 
scaling is different than that proposed here). 

Finally, we stress that the possibility of 
polarisation alignment in rotating turbulence (or a similar 
effect in MHD turbulence, discussed in \secref{sec_align_MHD})
does not undermine CB as a universal scaling 
conjecture --- what is revised in this version of CB turbulence is 
the assumption of two-dimensional isotropy in the perpendicular plane. 

\subsection{Polarisation alignment in MHD turbulence}
\label{sec_align_MHD}

The argument presented above is more or less analogous to the 
argument proposed by \citet{Boldyrev06} for MHD turbulence: 
he conjectured polarisation alignment between the perpendicular 
velocity and magnetic field fluctuations, which amounts 
to assuming that an Alfv\'en-wave polarisation is approximately 
preserved. The scalings he derived can be read off from 
\eqsref{E_rot_align}, \exref{aniso_rot_align} and \exref{Epar_rot_aniso}
by setting $v(\kperp)=v_A$ instead of $2\Omega/\kperp$ (note that 
\eqref{Epar_MHD} remains unchanged). 
Numerical simulations have confirmed both these scalings and
specifically the scale-dependent alignment between the fields 
(\citealt{Mason08,Boldyrev09}; see, however, 
\citealt{Beresnyak11}). It appears plausible that the 
$\kperp^{-2}$ spectra observed in simulations of forced rotating turbulence 
\citep{Mininni09,Thiele09} could be similarly explained by the 
polarisation alignment argument we have given here. However, a word 
of caution: most solar wind measurements 
of Alfv\'enic turbulence do not support the $\kperp^{-3/2}$ scaling
and incline rather towards $\kperp^{-5/3}$ 
\citep{Bale05,Horbury08,Sahraoui09,Wicks10,ChenMallet11}. 
Thus, it remains unclear whether polarisation alignment occurs in nature 
as well as in numerical boxes. 

We note in passing that 
polarisation alignment may also be responsible 
for the robust $\kperp^{-3/2}$ spectra reported in simulations 
of 2D MHD turbulence \citep{Biskamp89,Biskamp01}. There is 
no ``parallel'' Alfv\'enic propagation there, but assuming 
$k_x\gg k_y$, $u_x \ll u_y$, $\delta B_x \ll \delta B_y$, 
and $u_y \sim \delta B_y/\sqrt{4\pi\rho}$ 
still leads to a reduction of nonlinearity 
by $\theta\sim k_y/k_x$. This angle can again be assumed to scale
with the typical angular uncertainty at a given scale: 
$\theta\sim \delta B_y/\delta B_{\rm rms}$, where 
$\delta B_{\rm rms}$ is the rms amplitude of the magnetic 
fluctuations (i.e., the field at the outer scale). The argument 
is then the same as outlined in this section, leading to 
$E(\kperp)\sim (\eps v_A)^{1/2}\kperp^{-3/2}$, where 
$v_A = \delta B_{\rm rms}/\sqrt{4\pi\rho}\sim(\eps/k_0)^{1/3}$. 
The CB conjecture does not come in here because there is no 
parallel linear propagation. 
This argument highlights an aspect of MHD that is not analogous 
to the rotating case: a pure 2D state for the latter is simply 2D hydrodynamics, 
with no effect from the rotation (see the end of \secref{sec_inv_casc}),
whereas for MHD, setting $\dd/\dd z=0$ in \eqsand{eq_Phi_MHD}{eq_Psi_MHD} 
leaves velocities and magnetic fields still nonlinearly coupled via the 
Lorentz force. As we argued at the end of \secref{sec_gen}, however, a pure 2D 
state is not sustainable in a 3D world, so exact 2D MHD is an artificial situation.

\section{Stratified turbulence} 
\label{sec_strat}

Another hydrodynamic example 
where a critically balanced cascade should emerge 
is the stably vertically stratified turbulence, 
anisotropic with $\kpar\gg\kperp$, 
where $\kpar$ and $\kperp$ are the vertical and horizontal 
wave numbers, respectively \citep[see, e.g.,][]{Cambon01,Godeferd03,Laval03,Kaneda04}. 
The dispersion relation for (incompressible) gravity waves is 
$\omega = N\kperp/k\approx N\kperp/\kpar$, 
where $N$ is the Brunt--V\"ais\"al\"a frequency. 
Since the roles of $\kpar$ and $\kperp$ are reversed compared 
to the MHD and rotating turbulence, the arguments 
presented above have to be modified. 

First, the incompressibility $\vdel\cdot\vu=0$ now implies 
$\upar\sim(\kperp/\kpar)\uperp\ll\uperp$ (since $\kperp\ll\kpar$, 
it is no longer true that $\vdperp\cdot\vuperp=0$ 
as was the case for MHD and rotating turbulence). 
This implies that the nonlinear interaction time continues to be  
given by \eqref{tnl_gen}. If CB is assumed, 
the horizontal energy spectrum is, therefore, still 
Kolmogorov [\eqref{cb_gen}], while the relationship between 
the horizontal and vertical wave numbers is 
[analogous to \eqsand{aniso_MHD}{aniso_rot}]
\beq
\kperp \sim \eps N^{-3} \kpar^3 = l_O^2\kpar^3,
\label{aniso_str}
\eeq
where $l_O=\eps^{1/2}N^{-3/2}$ is called the \citet{Ozmidov92} scale. 
Using \eqref{aniso_str}, we can calculate the vertical 
energy spectrum corresponding to the horizontal spectrum \exref{cb_gen}
in a way analogous to the derivation of \eqref{Epar_rot}:
\beq
E(\kpar) \sim N^2\kpar^{-3},
\label{Epar_str}
\eeq
a spectrum previously proposed on dimensional grounds by \citet{Dewan97}
and by \citet{Billant01}. This argument is basically a reformulation in 
the CB language of the scaling hypothesis 
put forward by \citet{Lindborg06}, to whose paper we refer the reader 
for discussion and references on the atmospheric measurements 
and numerical simulations, which appear to be consistent 
with this theory \citep[see also][]{Kaneda04,Lindborg07}. 

The situation here is similar to rotating turbulence in that the anisotropy 
gets weaker as $\kpar$ increases. The cascade becomes isotropic 
at the Ozmidov scale, $\kperp\sim\kpar\sim l_O^{-1}$, where 
the local Froude number is $\uperp/l_O N\sim1$.  
At smaller scales, the turbulence cannot feel the mean gradient 
and becomes isotropic. Then both horizontal and vertical 
spectra are Kolmogorov, so there should be a spectral break at the 
Ozmidov scale for the vertical spectrum [transition from \exref{Epar_str}], 
but not the horizontal one, 
which is Kolmogorov already in the CB regime. 
The cascade path is similar to 
\figref{fig_rot} with $\kpar$ and $\kperp$ swapped 
and \eqref{aniso_str} used for the CB line.  

Despite these similarities, it must be acknowledged that because of the inversion 
of roles between $\kpar$ and $\kperp$, the case of stratified turbulence is perhaps 
somewhat special compared to the rather close analogy between MHD, plasma 
and rotating systems. In the latter cases, we had a two-dimensionally 
incompressible perpendicular turbulence and linear wave propagation in the 
1D parallel direction setting the correlations between distant perpendicular 
2D planes. In the case of stratified turbulence, the linear propagation is in 
the 2D horizontal plane, while the 1D vertical direction enters via 
wave dispersion and the nonlinearity ($\tnl^{-1}\sim\kperp\uperp\sim\kpar\upar$). 
Interestingly, the vertical spectrum \exref{Epar_str}, unlike its analogs 
\exref{Epar_rot} and \exref{Epar_MHD}, is independent of $\eps$. 
Such flux-independent spectra emerge quite commonly as a result of the 
break down of weak turbulence in 1D systems (e.g., \citealt{Newell08}; 
for discussions of universal flux-independent spectra see \citealt{Denissenko07} 
and \citealt[][p.~12]{Cardy08}). They are often associated with formation 
of singular structures (see \apref{ap_CS}), a phenomenon we do not expect to be 
a key player in MHD or rotating turbulence. Thus, the CB in stratified 
turbulence might not be the whole story and the matter deserves further study. 
One can also envision quite complicated regimes and multiple scale ranges 
emerging in systems that are both rotating and stratified, e.g., the Earth's 
atmosphere. 

\section{Conclusion} 
\label{sec_conc}

We have proposed that the critical balance 
of linear and nonlinear time scales, 
originally introduced for Alfv\'enic turbulence \citep{Higdon84,GS95}
and more recently, for other types of magnetised plasma turbulence 
\citep{Cho04,Tome}, should be used as a universal scaling conjecture 
for anisotropic turbulence in natural systems capable of supporting 
linear waves. While there are some indications 
that this idea works for stratified turbulence \citep{Lindborg06}, 
it has not been tested in rotating turbulence, for which we have 
proposed a novel energy cascade scenario and a set of verifiable predictions. 

In neutral fluids, the two examples we have considered --- rotating 
and stratified turbulence --- suggest that the critically balanced 
cascade provides a path from the strongly anisotropic 
fluctuations caused by the presence of an external field  
(mean rotation or gradient) to isotropic Kolmogorov turbulence 
inevitable at sufficiently small scales (cf.\ \citealt{Davidson06}). 
In that way, neutral fluids are different from conducting fluids and plasmas, 
where the presence of a mean magnetic field is felt at all scales 
and the anisotropy only gets stronger at smaller scales. 

%The idea that, between the linear and nonlinear physics, 
%nature ultimately refuses to discriminate in favour of one against 
%another appears to be a promising one. Indeed, in many wave 
%systems that do not quite fit the anisotropic type considered here, 
%similar situations seem to emerge: e.g., one might draw a parallel 
%between the critically balanced turbulence and the so-called generalised 
%Phillips spectrum associated with formation of whitecaps in surface
%water waves \citep{Newell08}.

We observe that one might draw parallels between the CB principle 
and the ``generalised Phillips spectra'' \citep{Newell08}
that are thought to emerge from the break down of weak turbulence 
in many wave systems, e.g., surface water waves \citep{Phillips58,Phillips85,Newell08}, 
Rossby waves \citep{Rhines75}, 
Kelvin waves in cryogenic turbulence \citep{Vinen00} 
and Bose-Einstein condensates \citep{Proment09}.
%See also discussions of universal flux-independent 
%spectra by \citet{Denissenko07} and \citet[][p.~12]{Cardy08}.
We note, however that some of these examples do not quite fit 
the anisotropic 3D type considered here and the break down of weak turbulence 
there may be related to formation of singular structures: 
this is discussed in more detail in \apref{ap_CS}. 

\begin{acknowledgments} 
We acknowledge important discussions with P.\ Berloff, S.\ Cowley, 
P.\ Davidson, K.\ Julien, J.\ McWilliams, F.\ Moisy and A.\ Newell.
We are also grateful to the numerous anonymous referees 
for raising a number of important issues, 
which led to significant expansion and improvement of the paper. 
This work was supported by STFC (A.A.S.) 
and the Leverhulme Trust Network for Magnetised Plasma Turbulence. 
\end{acknowledgments}

\appendix

\section{Reduced equations}
\label{ap_reduced}

A remarkable simplification of the underlying dynamical equations 
for turbulent fluctuations can be achieved by systematically 
taking into account their anisotropy. Here again CB
serves as a guiding principle, but this time as an ordering 
assumption: the expectation that the linear and nonlinear time 
scales would be comparable tells one how to order the fluctuation 
amplitudes with the expansion parameter $\epsilon=\kpar/\kperp$. 
This leads to reduced systems of equations, which are often more 
transparent physically and require less computational power to 
simulate numerically. An additional advantage of simulating reduced 
equations is that the transition to the asymptotic anisotropic regime 
is carried out analytically and so does not eat up resolution. 

Three well known examples of such reduced systems 
are the Reduced MHD (RMHD) equations for the 
Alfv\'enic turbulence (reproduced in \secref{ap_RMHD})
Electron Reduced MHD (ERMHD) equations 
for the turbulence of kinetic Alfv\'en waves at sub-Larmor scales
and Hall Reduced MHD (HRMHD) equations for 
Alfv\'enic turbulence in cold-ion plasmas  
(all three are derived under the CB ordering in \citealt{Tome}; 
the RMHD has been known since \citealt{Strauss76}; 
equations mathematically similar to ERMHD have been used to describe 
whistler turbulence in plasmas by, e.g., \citealt{Biskamp99}; 
HRMHD has been studied by many authors, e.g., \citealt{Gomez08}). 
A kinetic reduced system that emerges 
from the same principles is gyrokinetics \citep{Frieman82,Howes06}, 
a general description of magnetised plasma turbulence of which RMHD,  
ERMHD and HRMHD are particular limits. All of these equations have 
been successfully simulated numerically \citep{Perez08,Cho09,Gomez08,Howes08}; 
and in the case of RMHD and ERMHD, the results have explicitly been shown 
to be asymptotically consistent with simulations of the unreduced equations. 

Here we explain the procedure for deriving a reduced system 
on the example of rotating turbulence, showing again its very close 
resemblance to magnetised fluid systems and providing concrete justification 
for some of the assumptions made in the main text. 

\subsection{Reduced rotating hydrodynamics}

Our starting point is the Navier-Stokes equation for an incompressible 
fluid of viscosity $\nu$ and density $\rho=1$ rotating at the rate 
$\vOmega = \Omega\vz$ ($z$ is the rotation axis):
\beq
\frac{\dd\vu}{\dd t} + \vu\cdot\vdel\vu + 2\vOmega\times\vu 
= -\vdel p + \nu\nabla^2\vu,
\label{NSEq}
\eeq
where $p$ is pressure determined by $\vdel\cdot\vu=0$.
This equation supports inertial waves with frequency 
$\omega = \pm2\Omega \kpar/k$ and corresponding eigenfunctions
$\vu = (\pm ik_z/k, 1, \mp ik_x/k) u_y$, where 
$\vk=(k_x,0,k_z)$ without loss of generality. 
Note that for an inertial wave, the perturbed vorticity 
is aligned with velocity, $\vvort=\mp k\vu$, and so 
a monochromatic inertial wave is an exact nonlinear 
solution of \eqref{NSEq} 
(because $\vu\cdot\vdel\vu = \vvort\times\vu + \vdel u^2/2$; 
the gradient part can be absorbed into pressure).  

In the anisotropic regime (low Rossby numbers), 
\eqref{NSEq} is expanded in a small parameter $\epsilon=\kpar/\kperp$. 
Using the CB as an ordering prescription, we order 
the time scale of the fluctuations as 
$\omega\sim\epsilon\Omega\sim\vuperp\cdot\vdperp$. 
We also order $\upar\sim\uperp$ guided by the relationship 
between them in an inertial wave.

To lowest order in $\epsilon$, $\vdel\cdot\vu = \vdperp\cdot\vuperp = 0$, 
so the perpendicular motions are individually incompressible 
and can be represented by a stream function: $\vuperp=\vz\times\vdperp\Phi$. 
%Thus, the velocity field can be represented as $\vu = \vz\times\vdperp\Phi + \vz\upar$.
In the next order, the incompressibility condition allows us to find the 
divergence of the second-order correction to $\vuperp$ (to be useful shortly):
\beq
\label{inc2}
\vdel\cdot\vu = \vdperp\cdot\vuperp^{(2)} + \frac{\dd\upar}{\dd z} = 0. 
\eeq
The perpendicular part of \eqref{NSEq} is (keeping two lowest orders)
\beq
\label{NS_perp}
\frac{\dd\vuperp}{\dd t} + \vuperp\cdot\vdperp\vuperp -\nu\dperp^2\vuperp = 
-2\Omega\vz\times\vuperp -\vdperp p.
\eeq
In the lowest order, the left-hand side disappears and so 
right-hand side must vanish too. This gives $p=2\Omega\Phi$. 
Now taking the perpendicular curl ($\vdperp\times$) of \eqref{NS_perp}, 
we get
\beq
\frac{\dd}{\dd t}\dperp^2\Phi + \lt\{\Phi,\dperp^2\Phi\rt\} 
= 2\Omega\,\frac{\dd\upar}{\dd z} + \nu\dperp^4\Phi,
\label{Phi_eq}
\eeq
where $\{f,g\} = \vz\cdot(\vdperp f\times\vdperp g)$ 
and we have used \eqref{inc2} to express 
$\vdperp\times(2\Omega\vz\times\vuperp^{(2)}) = 2\Omega\vz\vdperp\cdot\vuperp^{(2)}$.
Finally, taking the parallel part of \eqref{NSEq} to lowest order 
and using $p=2\Omega\Phi$, we get
\beq
\frac{\dd\upar}{\dd t} + \lt\{\Phi,\upar\rt\} 
= -2\Omega\,\frac{\dd\Phi}{\dd z} + \nu\dperp^2\upar.
\label{upar_eq}
\eeq

\Eqsand{Phi_eq}{upar_eq} are the desired reduced system, 
which we will refer to as Reduced Rotating Hydrodynamics (RRHD). 
Up to notational differences, they are equivalent to the reduced 
system of \citet{Julien98} (derived under slightly differently 
formulated assumptions; see also \citealt{Julien07} and references 
therein for a uniform mathematical discussion of RRHD and 
reduced models generally). Let us itemise some of the properties 
of these equations:
\begin{enumerate}

\item they support inertial waves with $\omega = \pm2\Omega\kpar/\kperp$ 
and corresponding eigenfunctions $\upar=\pm\kperp\Phi$ (so, the wave 
is circularly polarised: as it propagates along $\vz$, the velocity vector 
rotates in the plane perpendicular to $\vkperp$); inertial-wave packets 
with fixed $\kperp$ and arbitrary amplitude 
are exact nonlinear solutions;

\item the velocity is $\vu=\vz\times\vdperp\Phi + \vz\upar$ 
and the (perturbed) vorticity $\vvort = -\vz\times\vdperp\upar + \vz\dperp^2\Phi$; 
therefore, the RRHD equations can be written (omitting viscosity)
\beq
\frac{\dd}{\dd t}\dperp^2\Phi + \lt\{\Phi,\dperp^2\Phi\rt\} 
= (\vvort + 2\vOmega)\cdot\vdel\upar,\qquad
\frac{\dd\upar}{\dd t} = -(\vvort+2\vOmega)\cdot\vdel\Phi,
\eeq
thus, besides the 2D self-advection of the perpendicular 
velocity, the basic physical process is propagation of inertial 
waves along the total vorticity lines (recall \secref{sec_aniso});

\item the 3D nature of the turbulence is enforced via {\em linear}
propagation terms (recall \secref{sec_gen}); when they are present, 
the system conserves one invariant, kinetic energy 
$\int d^3\vr\,\bl(|\vdperp\Phi|^2+\upar^2\br)/2$ and should have a direct 
cascade; when $\dd/\dd z=0$, there are three invariants: 
perpendicular kinetic energy $\int d^3\vr|\vdperp\Phi|^2/2$ 
(inverse cascade; recall \secref{sec_inv_casc}), 
enstrophy $\int d^3\vr\,|\dperp^2\Phi|^2/2$ (direct cascade), 
and parallel kinetic energy $\int d^3\vr\,\upar^2/2$ (direct cascade 
of a passive quantity);

\item while the RRHD equations were derived under the CB ordering, they 
allow both the weak and the strong turbulence regimes 
and so can track the transition from the former to the latter
(\secref{sec_unaligned}); 
they also remain valid if polarisation alignment occurs (\secref{sec_align})
and so can be used to measure and study it.

\item when nondimensionalising the RRHD equations, one can rescale
the parallel and perpendicular distances separately
(subject to appropriate rescaling of the amplitudes), 
i.e., the aspect ratio of the ``box'' is formally infinite --- this 
is because the anisotropy of the fluctuations was the basis for 
the asymptotic expansion that led to RRHD;
note that the scaling arguments of \secref{sec_rot} imply that 
the anisotropy of rotating turbulence diminishes with scale, 
so \eqsand{Phi_eq}{upar_eq} will produce solutions that, at 
sufficiently small scales, violate the ordering assumptions
behind the equations --- this should be 
manifested by the development of ever finer parallel structure 
(see \secref{sec_aniso}). 

\end{enumerate}

\subsection{Reduced magnetohydrodynamics}
\label{ap_RMHD}

Finally, for comparison, let us give the RMHD equations for 
the Alfv\'enic turbulence \citep[derived from MHD in exactly 
the same way as RRHD was from \eqref{NSEq}; see, e.g.,][section 2]{Tome}
\bea
\label{eq_Phi_MHD}
\frac{\dd}{\dd t}\dperp^2\Phi + \lt\{\Phi,\dperp^2\Phi\rt\} 
&=& v_A\frac{\dd}{\dd z}\dperp^2\Psi + \lt\{\Psi,\dperp^2\Psi\rt\}
= v_A\vb\cdot\vdel\dperp^2\Psi,\\
\label{eq_Psi_MHD}
\frac{\dd\Psi}{\dd t} &=& v_A\frac{\dd\Phi}{\dd z} + \lt\{\Psi,\Phi\rt\} 
= v_A\vb\cdot\vdel\Phi,
%\frac{\dd\upar}{\dd t} + \lt\{\Phi,\upar\rt\} &=& v_A^2\vb\cdot\vdel\frac{\dBpar}{B_0},\\
%\frac{\dd}{\dd t}\frac{\dBpar}{B_0} &=& \vb\cdot\vdel\upar,
\eea 
where $\Phi$ and $\Psi$ are the stream functions of the 
perpendicular velocity and magnetic fluctuations: 
$\vuperp=\vz\times\vdperp\Phi$, $\vBperp=\vz\times\vdperp\Psi$
(unlike in the rotating case, the parallel velocity and 
magnetic field fluctuations decouple and are passive 
with respect to the perpendicular ones; see \citealt{Tome}), 
and $\vb=\vz + \vBperp/B_0$ is the direction of the 
total magnetic field (along which the Alfv\'en waves 
propagate; recall \secref{sec_aniso}). 

The similarities with the equations for rotating turbulence 
are obvious, but there are also differences 
originating from the physical differences between the 
inertial and Alfv\'en waves: the latter are non-dispersive 
($\omega = \pm \kpar v_A$), the perpendicular velocity 
and magnetic field fluctuations are linearly polarised 
and the eigenfunctions $\Phi = \pm\Psi$ represent 
exact nonlinear solutions for arbitrary-shaped wave packets
(the \citealt{Elsasser50} solutions). It is also worth pointing 
out that whereas the coupling of $\Phi$ to $\upar$ (perpendicular 
to parallel velocity) in \eqref{Phi_eq} is purely linear, 
the coupling of $\Phi$ to $\Psi$ (velocity to magnetic field) 
in \eqref{eq_Phi_MHD} is both linear and nonlinear (via the Lorentz force).

\section{Critical balance and coherent structures}
\label{ap_CS}

It might appear tempting to relate the CB principle to the emergence of coherent 
structures for the following two reasons. Firstly, some coherent structures arise due to
wave breaking, e.g., in the system of water surface gravity waves or internal  
gravity waves, and such a wave breaking occurs precisely when the nonlinearity
becomes of order of the linear contributions.
Secondly, in some well known examples of coherent structures, such as solitons
or shocks, the linear and the nonlinear terms are in balance.
%Thus, let us consider in detail the question about the relation between the CB and
%coherent structures.

First, consider the coherent structures that result from wave breaking.
Such structures are typically singular, e.g., the wave profile has a derivative discontinuity.
To be specific, consider the water-surface gravity waves, where the CB
approach, i.e., scale-by-scale balance of the linear and the nonlinear time scales,
gives the well-known \citet{Phillips58} spectrum. 
Its connection to singular wave breaking structures 
has been widely discussed since it was first suggested
by \citet{Kadomtsev65} and later adopted 
by \citet{Phillips85} himself (his original 1958 paper does not mention wave breaking).  
However, there is an uncertainty related to the geometry and the dimensionality
of the wave crests. \citet{Kuznetsov04} argued that the Phillips spectrum
should correspond to wave crests with singularity at isolated points (cone-like shapes)
whereas more realistic 1D crests would give a different spectral exponent. 
Furthermore, coherent structures of different strengths or sizes can
coexist and the resulting spectrum can depend on their distribution (e.g., in his
refinement of the original theory \citealt{Phillips85} introduced a distribution function
for the crest lengths per unit area of the water surface). 
Thus, there is no  obvious universal link between the CB state
and singular coherent structures of the wave-breaking type: there may be structures
with spectra different from the CB prediction, but one can also imagine a CB state 
without any singular structures at all.

Now let us turn to the nonsingular coherent structures.
The most basic of the relevant nonlinear models is the Korteweg--de Vries (KdV) equation,
$u_t + u u_x + \mu u_{xxx} =0,$ where $\mu$ is the dispersion coefficient. 
Does CB work for KdV turbulence? 
Naively, the idea might seem promising because, 
the KdV model predicts formation of solitons --- coherent
structures in which the nonlinear and the linear terms are 
balanced, in the spirit of CB. Balancing the nonlinearity and dispersion 
scale by scale (the second and the third terms in the KdV equation), 
we get $E(k)\sim k^3$. 
However, if the separations between the solitons are much greater than their
widths, then for the scales intermediate between the soliton width and the inter-soliton separation, 
the spectrum is that of a set of delta-functions, so $E(k)\sim\const$, 
which is very different from the CB prediction. 
This is because the interaction is nonlocal in scale, 
%as easily verified from the KdV equation by substituting the soliton solution, 
whereas CB assumes locality.
Note also, that this is a 1D dispersive system and 
the physical arguments in favour of CB in anisotropic wave-supporting 
environments presented in \secref{sec_gen} do not generalise to it.  

In conclusion, there does not seem to be a universal link between the CB 
and coherent structures. In some systems there may be coherent structures but not 
CB because the latter assumes locality, which is not automatically guaranteed. 
Even if both coherent structures and a CB spectrum are present, the former need not 
be the cause of the latter as coherent structures might occupy a negligible volume. 
Finally, we reiterate that the physical argument for the formation of a CB state in MHD, 
which, as we showed above, may similarly be applied to the rotating turbulence, does not 
invoke either wave breaking or coherent structures but is rather based on 
local nonlinear energy transfer and anisotropic spatial decorrelation arguments 
(\secref{sec_gen}). As discussed at the end of \secref{sec_strat}, the case of stratified 
turbulence is more ambiguous because there a CB-based argument leads to a flux-independent 
vertical spectrum \exref{Epar_str} reminiscent of the Philips-type spectra 
produced by wave breaking. Whether this is a useful hint about the nature of 
stratified turbulence is clearly an intriguing question for future investigation. 

\bibliographystyle{jfm}
\bibliography{ns_JFM}
\end{document}